\documentclass[aps,onecolumn,superscriptaddress,groupedaddress,nofootinbib]{revtex4}
\usepackage{graphicx}% Include figure filesa
\usepackage{dcolumn}% Aligntable columns on decimal point
\usepackage{bm}
\usepackage{mathrsfs}% bold math
\usepackage{hyperref}% add hypertext capabilities
\usepackage{amsmath}
\usepackage{amssymb}
\usepackage{bm}
\usepackage{color}
\usepackage{float}
\usepackage{dcolumn}
\usepackage{multirow}
\usepackage{changepage}
\usepackage{enumerate}
\usepackage{setspace}
\hypersetup{pdftex,colorlinks=true,linkcolor=blue,citecolor=red,menucolor=black,urlcolor=blue,filecolor=blue}

\newcommand{\mev}{\textrm{ MeV}}

\begin{document}

\title{The $D^+_s \to K^+ \pi^+ \pi^-$ reaction and the scalar $f_0(500)$, $f_0(980)$ and $K^*_0 (700)$ resonances }

\author{L. R. Dai}
\email[]{dailianrong@zjhu.edu.cn}
\affiliation{School of Science, Huzhou University, Huzhou 313000, Zhejiang, China}
\affiliation{Departamento de F\'{\i}sica Te\'orica and IFIC, Centro Mixto Universidad de Valencia-CSIC Institutos de Investigaci\'on de Paterna, Aptdo.22085, 46071 Valencia, Spain}

\author{E. Oset}
\email[]{oset@ific.uv.es}
\affiliation{Departamento de F\'{\i}sica Te\'orica and IFIC, Centro Mixto Universidad de Valencia-CSIC Institutos de Investigaci\'on de Paterna, Aptdo.22085, 46071 Valencia, Spain}

\begin{abstract}
  We develop a model to reproduce the mass distributions of pairs of mesons in the Cabibbo-suppressed
 $D^+_s \to K^+ \pi^+ \pi^-$ decay. The largest contributions to the process comes from the  $D^+_s \to K^+ \rho^0$ and $D^+_s \to K^{*0} \pi^+$ decay modes,
 but the  $D^+_s \to  K^*_0(1430) \pi^+$  and $D^+_s \to K^+  f_0(1370)$ modes also play a moderate role and all of them are introduced
empirically. Instead, the contribution of the $f_0(500)$, $f_0(980)$ and $K^*_0(700)$ resonances is introduced dynamically
by looking at the decay modes at the quark level, hadronizing $q \bar{q}$
pairs to give two mesons, and allowing these mesons to interact to finally produce the $K^+ \pi^+ \pi^-$  final state.
These last three modes are correlated by means of only one parameter. We obtain a fair reproduction of the experimental data for the three mass distributions as well as the relative weight of the three light scalar mesons, which we see as further support for the nature of these states as dynamically generated from the interaction of pseudoscalar mesons.

\end{abstract}

\maketitle

\section{Introduction}
 The hadronic weak decays of $D, D_s$ mesons are an excellent source of information on the interaction of hadrons \cite{petrov,myreview}. In particular, decays of $D, D_s$ into three mesons allow one to study the interaction of pairs of particles at different invariant
 masses and observe hadronic resonances. One case which has attracted much attention is the decay with one kaon in the final state,
 $D \to K \pi \pi (\eta)$ \cite{kaminsky,xiedai,kubis,toledo,mousssallam}, and the simultaneous study of the  $D^+ \to K^- \pi^+ \pi^+$  and $D^+ \to K_S^0 \pi^0 \pi^+$ reactions
 is done in \cite{newkubis}. Related work on $D \to K K \pi$ is also
 addressed in \cite{enwang,gengxie,sunxiao,wanggeng},  $D^+_s \to \pi^+ K^0_S K^0_S$ in \cite{dai}
  and $D \to K K K$  is also addressed in \cite{patricia,luisroca}. In the present work we study the singly Cabibbo-suppressed $D_s \to K^+ \pi^+ \pi^-$ decay. The reaction has been measured in \cite{focus} and more recently,
  with better statistics, in \cite{besexpe}. In addition to the dominant mode $D^+_s \to K^+\rho, \rho \to \pi^+ \pi^- $ and
  $D^+_s \to  K^*(892)^0 \pi^+, K^*(892)^0 \to K^+  \pi^-$, the experiment finds traces of the $f_0(500)$, $f_0(980)$ and $f_0(1370)$ resonances. No theoretical work on this particular channel  is available  to the best of our knowledge, and  we wish to address this problem here. The procedure followed follows the line of related work in which the dominant weak decay modes at the quark model are investigated and hadronization of quark pairs is considered to convert the first step weak decay into the production of three mesons.
After this first step, the different meson pairs are allowed to interact to lead to the final  observed channel \cite{xiedai,enwang,gengxie,sunxiao,wanggeng}. In  the process of interaction some resonances are generated, and in the light meson sector this task is undertaken using the chiral unitary approach \cite{ollerramos}. We shall see that in this process we generate the  $f_0(500), f_0(980)$ and $K^*_0 (700)$ scalar
resonances from the $\pi\pi, K\bar{K}$ and $K\eta$, $K\pi$ interaction respectively, providing support to the dynamical generation of these resonances.

\section{Formalism}

\begin{figure}[h!]
\centering
\includegraphics[scale=.7]{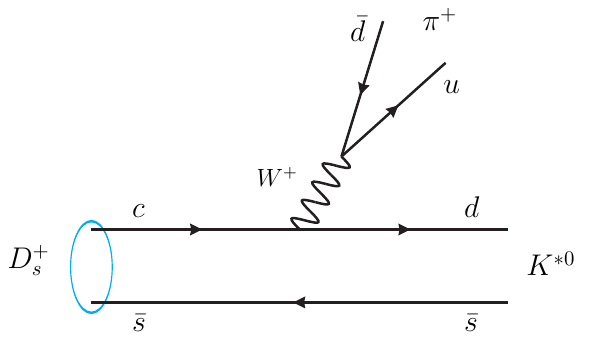}
\caption{Mechanism for production of  $\pi^+ K^{*0}$ in $D^+_s$ decay with external emission}
\label{fig:1}
\end{figure}

\begin{figure}[h!]
\centering
\includegraphics[scale=.8]{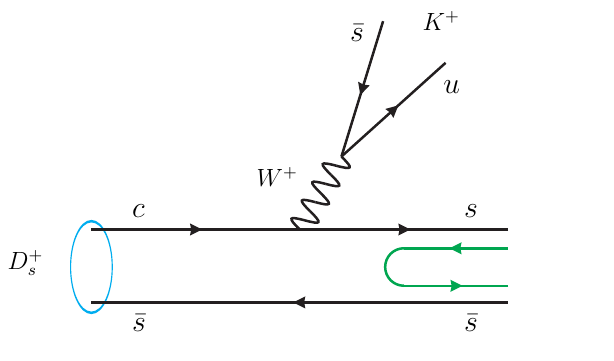}
\caption{$D^+_s \to K^+ s \bar{s}$ with external emission and $s \bar{s}$  hadronization}
\label{fig:2}
\end{figure}

\begin{figure}[h!]
\centering
\includegraphics[scale=.8]{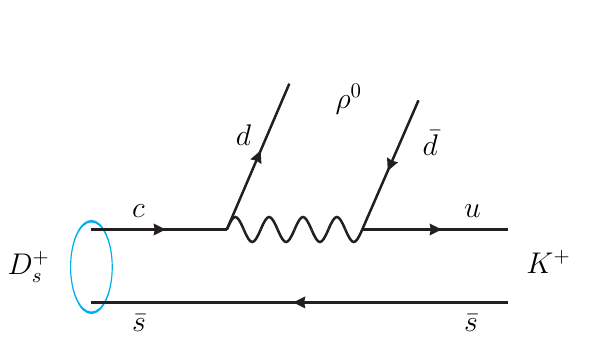}
\caption{Mechanism for $D^+_s \to \rho K^+$ with internal emission}
\label{fig:3}
\end{figure}

\begin{figure}[h!]
\centering
\includegraphics[scale=.8]{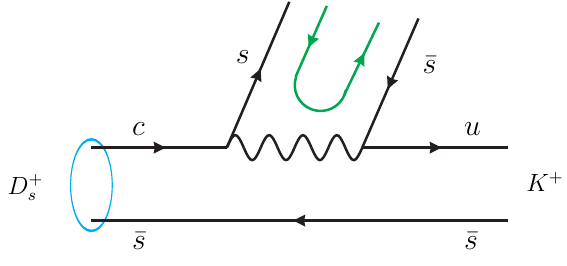}
\caption{$D^+_s \to K^+ s \bar{s} $ with internal emission followed  by $s \bar{s}$ hadronization}
\label{fig:4}
\end{figure}

\begin{figure}[h!]
\centering
\includegraphics[scale=.8]{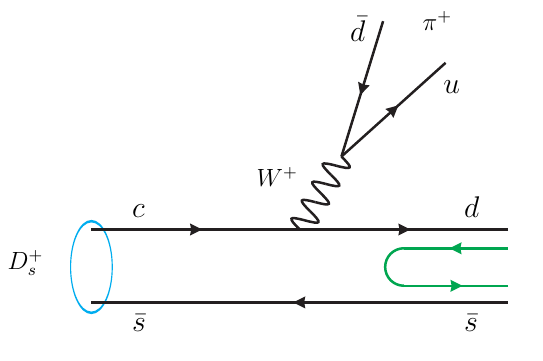}
\caption{$D^+_s \to  \pi^+ d \bar{s}$ with external emission and  $d \bar{s}$ hadronization}
\label{fig:5}
\end{figure}

\begin{figure}[h!]
\centering
\includegraphics[scale=.8]{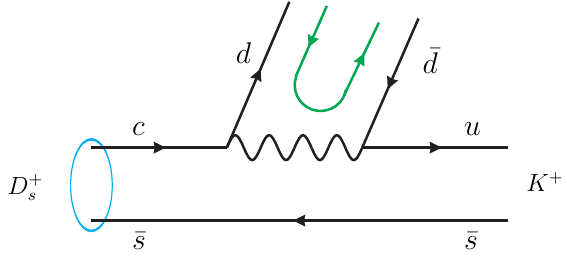}
\caption{$D^+_s \to K^+ d \bar{d} $ with internal emission and  $d \bar{d}$ hadronization}
\label{fig:6}
\end{figure}

In the $D^+_s \to K^+ \pi^+ \pi^-$ reaction one can guess that both the $K^{*0}$ decaying to $K^+  \pi^-$ and the $\rho^0$ decaying to
$\pi^+ \pi^-$ are formed. We shall see that this is the case. To understand the process we look at the $D^+_s$ decay at the quark level which shows that the process proceeds via a Cabibbo-suppressed mechanism. Instead of having the Cabibbo-allowed $W^+ \to c\bar{s}, u\bar{d}$ vertices,
we have now $W^+ \to c\bar{d}, u\bar{s}$ as one can see in Figs.~\ref{fig:1}-\ref{fig:6}.

The fact that we have two ways involving $\sin\theta_c$ from the $W^+  \to c \bar{d}$ and  $W^+  \to u \bar{s}$ vertices is responsible for the
many diagrams that we have obtained leading to this singly  Cabibbo-suppressed decay mode. We shall give a weight to the
different diagrams according to the following scheme:
\begin{itemize}
  \item[1)] weight $\alpha$ for $K^{*0}$ production
  \item[2)] weight $\alpha\,h$, the $h$ factor accounting for the mechanism of hadronization
  \item[3)] weight $\gamma$ for $\rho^{0}$ production
  \item[4)] weight $\gamma\,h$ since it involves an extra hadronization as in the case of 2)
  \item[5)] weight $\alpha\,h$ since it has  the same topology as the case  of 2)
  \item[6)] weight $\gamma\,h$  since it involves an extra  hadronization with respect to case   of 3)
\end{itemize}

Next we proceed to look in detail into the different  hadronization processes.
In Figs.~\ref{fig:2}, \ref{fig:4} we have the hadronization  of the $s\bar{s}$ component and we add a $\bar{q}q$ pair with the quantum numbers
of the vaccum. By writing the $q_i \bar{q}_j$ matrix of SU(3) in terms of the pseudoscalar mesons we have

\begin{eqnarray}\label{eq:p}
q\bar{q}  \to P=  \left(
  \begin{array}{ccc}
 \frac{\pi^0 }{\sqrt{2}}+ \frac{\eta}{\sqrt{3}}  & \pi^+ & K^+ \\[2mm]
   \pi^- & -\frac{\pi^0 }{\sqrt{2}}+ \frac{\eta}{\sqrt{3}}  & K^0 \\[2mm]
  K^- & \bar{K}^0 & -\frac{\eta}{\sqrt{3}}\\
     \end{array}
    \right)\,,
\end{eqnarray}
where we have taken the standard  $\eta$ and $\eta'$ mixing of Ref.~\cite{bramon} and neglected the $\eta'$  which does not play a role
in the generation  of the resonances that we shall discuss. Then
\begin{eqnarray}\label{eq:ssb}
s\bar{s} &\to& \sum_i s\, \bar q_i q_i \, \bar s  =\sum_i P_{3i} P_{i3} = (P^2)_{33} \nonumber\\
&=&K^- K^+ + \bar{K}^0 K^0+ \frac{1}{3} \eta \eta
\end{eqnarray}
In Fig.~\ref{fig:5} we have the hadronization of $d\bar{s}$ as
\begin{eqnarray}\label{eq:dsb}
d\bar{s} &\to& \sum_i d\, \bar q_i q_i \, \bar s = \sum_i P_{2i} P_{i3} = (P^2)_{23}\nonumber\\
&=&\pi^- K^+  - \frac{1}{\sqrt{2}} \pi^0 K^0
\end{eqnarray}

In Fig.~\ref{fig:6} we have the hadronization of $d\bar{d}$ as
\begin{eqnarray}\label{eq:ddb}
d\bar{d} &\to& \sum_i d\, \bar q_i q_i \, \bar d = (P^2)_{22} \nonumber\\
&=&\pi^- \pi^+  +  \frac{\pi^0 \pi^0}{\sqrt{2}} + \frac{\eta \eta}{3} -\frac{2}{\sqrt{6}} \pi^0 \eta +  K^0 \bar{K}^0
\end{eqnarray}

The 4) and 6) cases correspond to the same topology and have the same weight and can be summed into

\begin{eqnarray}\label{eq:plus46}
 4)+6) &\to&  (P^2)_{33} +(P^2)_{22}  \nonumber\\
&=&\pi^+  \pi^-  + \frac{\pi^0 \pi^0}{\sqrt{2}} + \frac{2}{3} \eta \eta +  K^+ K^-
 + 2 K^0 \bar{K}^0 -\sqrt{\frac{2}{3}}  \pi^0 \eta ~~
\end{eqnarray}

\begin{figure}[h!]
\centering
\includegraphics[scale=.8]{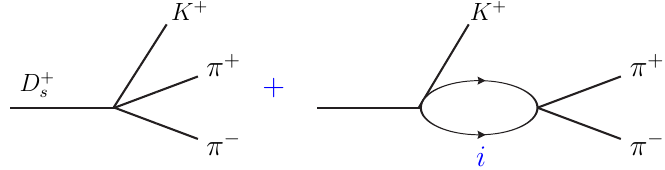}
\caption{Direct $K^+ \pi^- \pi^+ $ production  (tree level) and production through intermediate states,
 $i=\pi^+ \pi^-$, $\pi^0 \pi^0$, $\eta \eta$,  $\pi^0 \eta$,  $K^+ K^-$,  $K^0 \bar{K}^0$ in general.  }
\label{fig:7}
\end{figure}

We  can see that in Fig.~\ref{fig:6} we already obtain  $K^+ \pi^- \pi^+ $ at the tree level, but we also get other intermediate states
that upon rescattering lead to the same state, as depicted in  Fig.~\ref{fig:7}

Given Eqs.~\eqref{eq:ssb}, \eqref{eq:dsb}, \eqref{eq:ddb}, \eqref{eq:plus46}, we  can write the production matrix
for each mechanism of Figs.~\ref{fig:1} to \ref{fig:6}.
\begin{eqnarray} \label{eq:t2}
t^{(2)}=\sum_i \alpha \, h \, W_i \,G_i (M_{\rm inv}, \pi\pi) \, t_{i,\pi^+ \pi^-} (M_{\rm inv}, \pi\pi)
\end{eqnarray}
 with $i=K^+ K^-, K^0 \bar{K}^0, \eta\eta$, where the weights $W_i$ are given by means of  Eq.~\eqref{eq:ssb} as
 \begin{eqnarray}
 W_{K^+ K^-}=1\,,    W_{K^0 \bar{K}^0}=1 \,,  W_{\eta\eta}=\frac{2}{3} \frac{1}{\sqrt{2}}    \nonumber
\end{eqnarray}

In the case of the two identical particles $\eta\eta$ we have considered the factor $2$ for the two particles and $\frac{1}{\sqrt{2}}$
 because we work with the unitary normalization where the state is normalized as $\frac{1}{\sqrt{2}}\eta\eta $  to avoid double counting in the $G$ loop function. In  Eq.~\eqref{eq:t2}  $G (M_{\rm inv})$ is the diagonal loop function of two intermediate pseudoscalar mesons, which
 we regularize with a cut off of $q_{\rm max}=600\mev$, and $t_{i,j}$ are the transition scattering matrices for the six pseudoscalar pairs,
 $\pi^+ \pi^-,\pi^0 \pi^0,K^+ K^-,K^0 \bar{K}^0,\eta\eta,\pi^0 \eta$ obtained in a coupled channel formalism as
 \begin{eqnarray}\label{eq:bs}
 t=[1-V G]^{-1} V
\end{eqnarray}
with the transition potentials $V_{ij}$ obtained from \cite{jiangliang}. For the $ \pi^- K^+$ and   $\pi^0 K^0$ interaction
we use Eq.~\eqref{eq:bs} with the coupled channels $ \pi^- K^+, \pi^0 K^0, \eta K^0$ with the  transition potentials of Ref. \cite{xiedai,toledo}.

Similarly, we obtain
 \begin{eqnarray} \label{eq:tp6}
t^{(4+6)}= \gamma  \, h\, \left\{1+ \sum_i  W'_i  \,G_i (M_{\rm inv}, \pi\pi)  \, t_{i,\pi^+ \pi^-} (M_{\rm inv}, \pi\pi) \right\}~
\end{eqnarray}
with  $i=\pi^+ \pi^-,\pi^0 \pi^0,K^+ K^-,K^0 \bar{K}^0,\eta\eta,\pi^0 \eta$ and
 \begin{eqnarray}
 W^{\prime}_{\pi^+ \pi^-}=1 \,,  W'_{\pi^0 \pi^0}=2\, \frac{1}{2}\, \frac{1}{\sqrt{2}} \,,   W^{\prime}_{K^+ K^-}=1  \nonumber\\
  W^{\prime}_{K^0 \bar{K}^0}=2\,,  W'_{\eta\eta}= \frac{2}{3} \frac{1}{\sqrt{2}} \,2 \,,   W'_{\pi^0  \eta}= -\sqrt{\frac{2}{3}}
\end{eqnarray}
\begin{eqnarray} \label{eq:t5}
t^{(5)}=\alpha \,h \, \left\{1+\sum_i \widetilde{W}_i \,G_i (M_{\rm inv},\pi^- K^+) \,t_{i,\pi^- K^+}(M_{\rm inv},\pi^- K^+)\right\}~
\end{eqnarray}
with $i=\pi^- K^+, \pi^0 K^0$ and
\begin{eqnarray}
\widetilde{W}_{\pi^- K^+}=1 \,,  \widetilde{W}_{\pi^0 K^0}=-\frac{1}{\sqrt{2}}
\end{eqnarray}
Note that in Eqs.~\eqref{eq:tp6}, \eqref{eq:t5} we have the term $1$ in the amplitude, which correspond to the tree level $K^+ \pi^+ \pi^- $
production. This term is absent in Eq.~\eqref{eq:t2} since the primary production does not contain $K^+ \pi^+ \pi^- $.

\subsection{vector production}
\begin{figure}[h!]
\centering
\includegraphics[scale=.8]{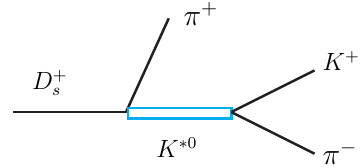}
\caption{Mechanism  for $D^+_s \to \pi^+ K^{*0}, K^{*0} \to K^+ \pi^- $}
\label{fig:8}
\end{figure}

\begin{figure}[h!]
\centering
\includegraphics[scale=.8]{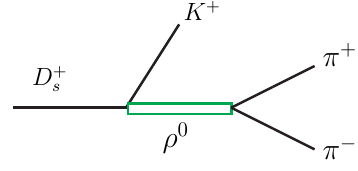}
\caption{Mechanism  for $D^+_s \to K^+ \rho^0, \rho^0 \to \pi^+ \pi^- $}
\label{fig:9}
\end{figure}

We look now to the mechanisms of Figs.~\ref{fig:1} and \ref{fig:3} for $K^{*0}$ and $\rho^0$  production respectively. We show these
processes in Figs.~\ref{fig:8}, \ref{fig:9} respectively, including the $K^{*0}$ and $\rho^0$ decays.

In order to obtain the $K^{*0} \to K^+ \pi^-$ and $\rho^0 \to \pi^+ \pi^- $ vertices  we use the standard Lagrangian
\begin{eqnarray}\label{eq:L}
  {\cal{L}} &=& -i \,g \,\langle [P, \partial_\mu P] V^\mu\rangle \,,   \nonumber\\
  g &=& \frac{M_V}{2\,f}~(M_V=800 \mev,~f=93 \mev)
\end{eqnarray}
with $\langle \rangle$ indicating the SU(3) trace and  $V^\mu$ given by
\begin{eqnarray}\label{eq:V}
V^\mu = \left(
  \begin{array}{ccc}
 \frac{\rho^0 }{\sqrt{2}}+ \frac{\omega}{\sqrt{2}}  & \rho^+ & K^{*+} \\[2mm]
   \rho^- & -\frac{\rho^0 }{\sqrt{2}}+ \frac{\omega}{\sqrt{2}}  &K^{*0} \\[2mm]
  K^{*-} & \bar{K}^{*0} & \phi\\
     \end{array}
    \right)\,.
\end{eqnarray}

The vertex $D^+_s \to K^+ \rho^0$ has the same structure and we take
\begin{eqnarray}
V_{D^+_s \to K^+ \rho^0} \equiv \epsilon^\nu (P_{D_s}+ P_{\pi^+ })_\nu
\end{eqnarray}

Following the  lines detailed in Ref. \cite{toledo} we can write the amplitude in terms of the invariant masses $s_{12},s_{13},s_{23}$ for the
particles in the order $\pi^- (1),\pi^+ (2), K^+ (3) $ as
\begin{eqnarray}
t^{(1)} &=& \alpha \, g \frac{1}{s_{13}-m^2_{K^*}+i m_{K^*} \Gamma_{K^*}} \nonumber\\
&\times &\left\{-s_{23}+s_{12} + \frac{(m^2_{K^+}-m^2_{\pi^-})(m^2_{D_s}-m^2_{\pi^+})}{m^2_{K^*}}\right \} ~~~~~
\end{eqnarray}
and similarly
\begin{eqnarray}
t^{(3)}=\gamma \, g \, \sqrt{2} \frac{1}{s_{12}-m^2_{\rho}+i\, m_{\rho} \Gamma_{\rho}}
\left\{-s_{13}+s_{23}\right \}
\end{eqnarray}
and we use the relationship
\begin{eqnarray}
s_{12}+ s_{23}+s_{13}=m^2_{D_s}+m^2_{K^+}+m^2_{\pi^+}+m^2_{\pi^-}
\end{eqnarray}

\subsection{Higher mass scalar resonances}
Following the analysis of the experimental work \cite{besexpe}, we also allow the contribution of two
scalar resonances, the $f_0(1370)$ and $K^*_0(1430)$. These resonances do not come from pseudoscalar-pseudoscalar
interaction but, instead, they are obtained from vector-vector interaction, together with many other states with $J=0,1,2$ \cite{raquel,geng}. Yet, these are the two resonances which
are obtained with less precision in \cite{raquel,geng}, with $150-200\mev$ difference in the mass with respect to the experiment, hence here we do not try to obtain them  in the way we have dealt with the light
scalar resonances and introduce them empirically with weights as free parameters.

\begin{figure}[h!]
\centering
\includegraphics[scale=.8]{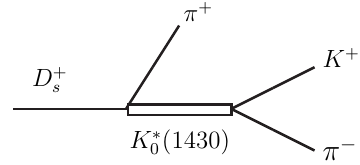}
\caption{Mechanism  for $D^+_s \to \pi^+ K^*_0(1430),   K^*_0(1430)\to  K^+  \pi^- $}
\label{fig:new1}
\end{figure}

\begin{figure}[h!]
\centering
\includegraphics[scale=.8]{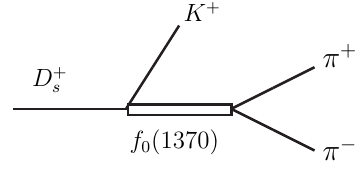}
\caption{Mechanism  for $D^+_s \to K^+ f_0(1370,  f_0(1370\to \pi^+ \pi^- $}
\label{fig:new2}
\end{figure}

The mechanisms for the production of these resonances are depicted in Figs.~\ref{fig:new1}, \ref{fig:new2},
and their amplitudes can be parameterized by means of
\begin{eqnarray}
t^{(7)}=\beta\,  \frac{m^2_{D_s}}{s_{13}-m^2_{K^*_0(1430)} + i\, m_{K^*_0(1430)} \Gamma_{K^*_0(1430)}}
\end{eqnarray}
for $K^*_0(1430)$ production and
\begin{eqnarray}
t^{(8)}=\delta \,  \frac{m^2_{D_s}}{s_{12}-m^2_{f_0(1370} + i\, m_{f_0(1370} \Gamma_{f_0(1370}}
\end{eqnarray}
for $f_0(1370)$  production, where the factor $m^2_{D_s}$ is introduced to have $\beta, \delta$ dimensionless.
We take  the masses and widths from the PDG \cite{pdg}, $M=1425\mev$, $\Gamma=270\mev$ for $K^*_0(1430)$ meson, and $M=1370\mev$, $\Gamma=350\mev$  for $f_0(1370)$ meson.

The sum of all contributions is  given by
\begin{eqnarray}
t=t^{(1)}+t^{(3)}+t^{(2)}+t^{(4+6)}+ t^{(5)}+ t^{(7)}+ t^{(8)}
\end{eqnarray}
and to get the mass distribution we use the PDG formula \cite{pdg}
\begin{eqnarray}\label{eq:pdg}
\frac{d^2\Gamma}{dm^2_{12} dm^2_{23}}=\frac{1}{(2\pi)^3} \,\frac{1}{32 M^3_{D_s}} |t|^2
\end{eqnarray}
where $m^2_{12}=s_{12}$, $m^2_{23}=s_{23}$ for $\pi^+ \pi^- $, $\pi^+ K^+ $ respectively. We integrate  Eq.~\eqref{eq:pdg}
 over $s_{23}$ with the limits of the PDG \cite{pdg} and obtain $d\Gamma/dm^2_{12}$. By cyclical permutation of the indices we easily obtain
$d\Gamma/dm^2_{13}$ and $d\Gamma/dm^2_{23}$.

\section{Results}
\begin{figure}[h!]
\centering
\includegraphics[scale=.8]{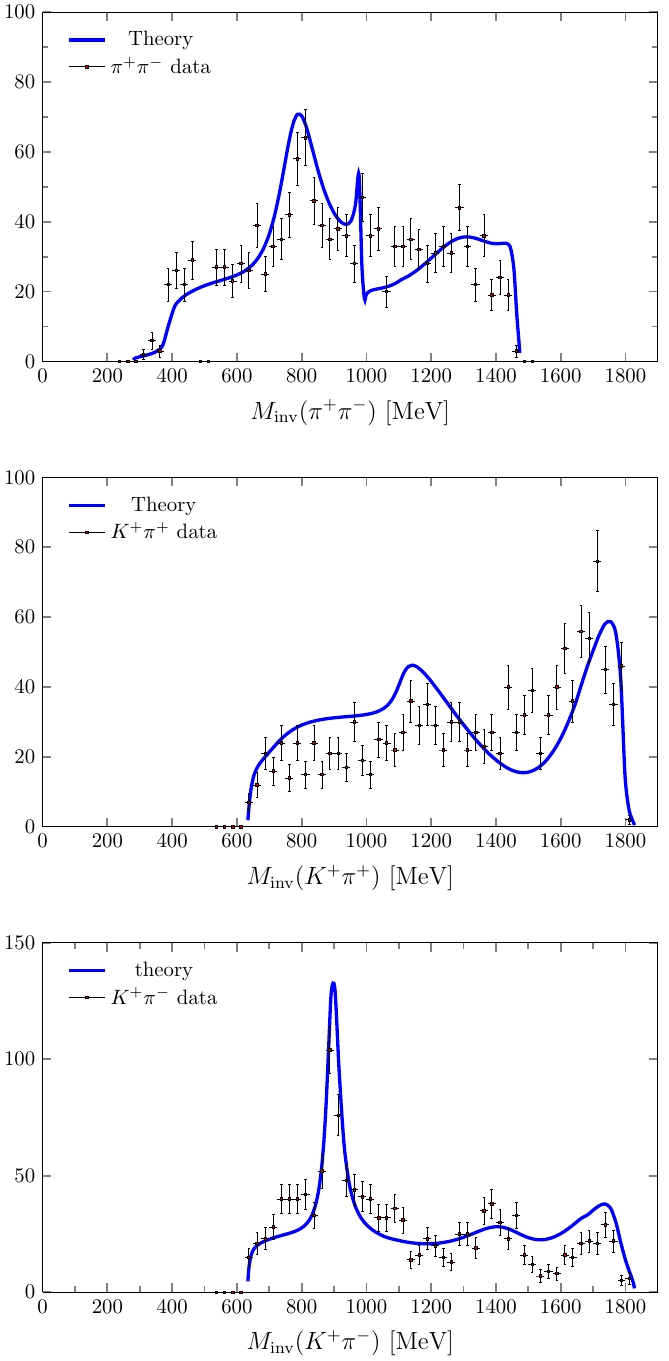}
\caption{Invariant mass distributions}
\label{fig:10}
\end{figure}

We conduct a best fit to the three invariant mass distributions of Ref. \cite{besexpe} and we get the values for the parameters
\begin{eqnarray}
\alpha=14.67\,,  \gamma=10.75\,, h=6.86\,,
\beta=-33.23\,, \delta=-58.84
\end{eqnarray}
The results for the mass distributions  are shown in Fig.~\ref{fig:10}.
The agreement with the data is fair and the $K^{*0}$, $\rho^0$ peaks are prominent in the reaction.

\begin{figure}[h!]
\centering
\includegraphics[scale=.8]{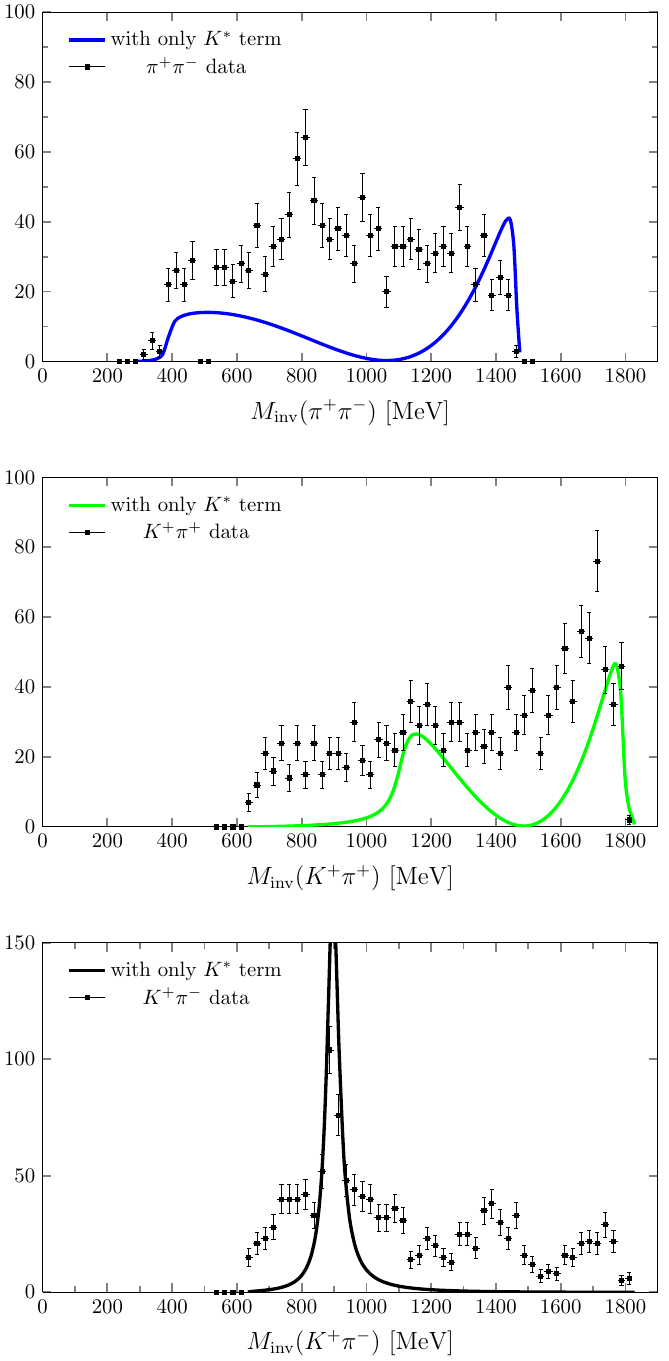}
\caption{Invariant mass distributions obtained with the  $K^{*0}$  term alone.}
\label{fig:11}
\end{figure}

\begin{figure}[h!]
\centering
\includegraphics[scale=.8]{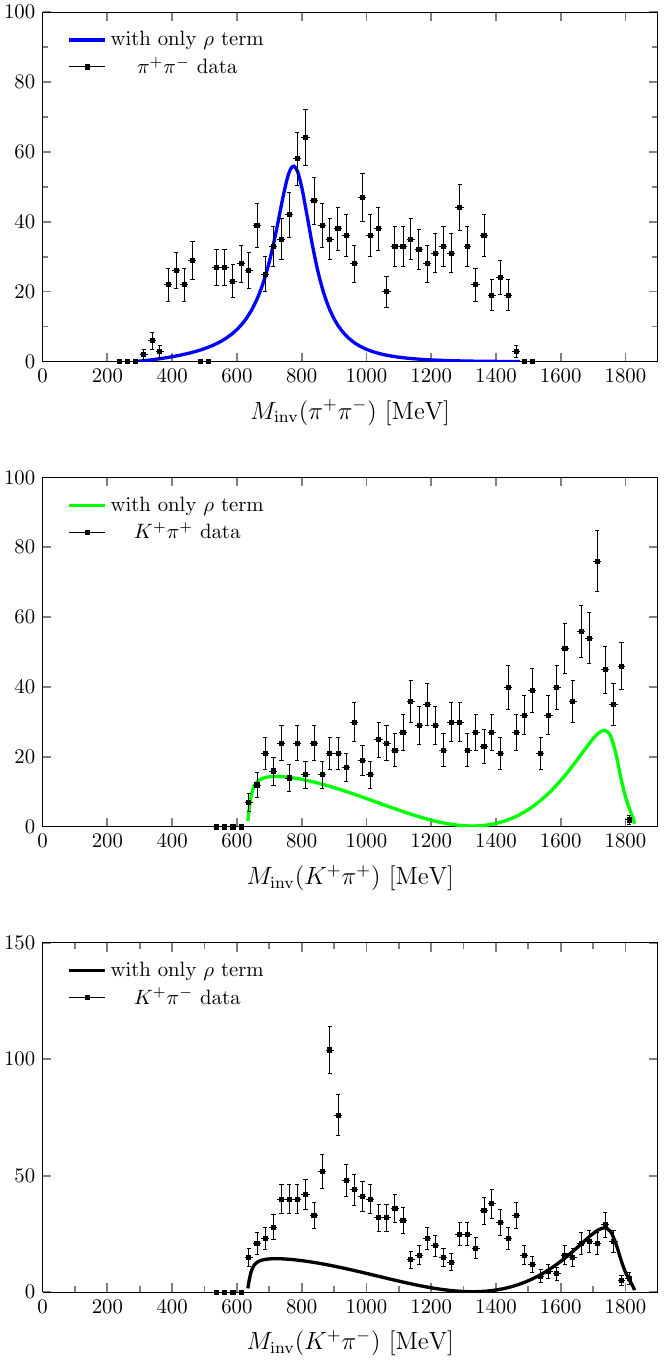}
\caption{Invariant mass distributions obtained with the   $\rho^0$   term alone.}
\label{fig:12}
\end{figure}

The $K^*_0(1430)$ contribution is observed as a soft peak in $K^+  \pi^- $ mass spectrum of Fig.~\ref{fig:10} around $1400\mev$
and the $f_0(1370)$, which has a very large width, shows up in the region around $1200-1400\mev$, where otherwise there would be strength  missing.
On the other hand, the $f_0(500)$, $f_0(980)$, $K^*_0 (700)$ have been introduced dynamically here, through the interaction of pseudoscalar pairs, and
one can see their contribution in the low energy part of the $M_{\rm inv} (\pi^+ \pi^-)$ spectrum of Fig.~\ref{fig:10}, the sharp peak around $980\mev$
in the same spectrum and the low energy part of the $K^+  \pi^- $ mass spectrum in the same  figure, respectively.

Technically, from the amplitude $t^{(2)}$, since the $K\bar{K}$, $\eta\eta$ come from the  $s\bar{s}$  hadronization  that has $I=0$, we can expect
to obtain a contribution from the  $f_0(980)$, which couples strongly to $K\bar{K}$ but weakly to $\pi \pi$, and to a minor extend
a contribution from the $f_0(500)$ which couples to $\eta\eta$ but not strongly. On the other hand, from $t^{(4+6)}$ we get
contribution both from $f_0(980)$ and $f_0(500)$  since now we have $\pi \pi$ intermediate  states which couple strongly to
$f_0(500)$. Furthermore, from  $t^{(5)}$ we get a contribution from the scalar $K^*_0 (700)$ resonance which couples to $K \pi$.

We should note that all these three resonance contributions have been included by means of a unique parameter, $h$, and the fair reproduction
 of the spectra obtained supports that these  contributions are indeed correlated and  our mechanism for production of these resonances produces a fair
 reproduction of their relative weight in these mass distributions.

It is interesting to see what distributions we obtain if we keep only the $K^{*0}$  or the  $\rho^0$  terms. This is shown in
Figs.~\ref{fig:11} and \ref{fig:12}. We observe in Fig.~\ref{fig:11} that much of the strength  in the $K^+  \pi^- $ mass distribution  outside the $K^{*0}$ peak is not accounted for.
On the other hand  it produces a two peak structure in the  $K^+  \pi^+ $ distribution and also in the $ \pi^+ \pi^-$ one.
These peaks are well known as reflections  in some channels of resonances in another channel and should not be confused with signals of a new resonance.
 In Fig.~\ref{fig:12} we repeat the exercise  putting only the contribution of the $\rho$. Once again, we show that much strength outside the $\rho$
 region is not accounted for and, similarly to the case of the $K^*$ resonance alone, the $\rho$ peak generate reflections with two peaks, both in the
 $K^+  \pi^+ $  and $K^+  \pi^- $ mass distributions.

\section{Conclusions}
   We have performed a fit to the three mass distributions of the
   $D^+_s \to K^+ \pi^+ \pi^-$   reaction in which we have introduced empirically the contributions of the main decay channels, $D^+_s \to K^+ \rho^0$ and $D^+_s \to K^{*0}  \pi^0$. In addition, we also introduce empirically two other contributions from channels of smaller relevance, the
   $D^+_s \to \pi^+ K^*_0(1430) $ and $D^+_s \to K^+ f_0(1370)$.
    The novelty of our approach is that we introduce the contribution from the $f_0(500)$, $f_0(980)$ and $K^{*}_0(700)$ resonances from the perspective that they are dynamically generated resonances,
     stemming the interaction of pseudoscalar mesons. For this purpose we look at the decay channels at the quark level, perform a hadronization of $q\bar{q}$ pairs to produce three pseudoscalar mesons in the final state, and allow these mesons to interact by pairs to produce the desired final state. In this way the three light scalar mesons are introduced dynamically and their contributions are correlated by means of just one free parameter.  We obtain a fair reproduction of the $\pi^+ \pi^-$, $K^+ \pi^-$ and
    $K^+ \pi^+$ mass distributions and the relative weight of the contribution of the light scalar mesons also agrees with the measured spectra. We see these features as extra support for the dynamically generated origin of these resonances, stemming from the interaction of pseudoscalar mesons, which in the present case is considered using the chiral unitary approach.

\section*{Acknowledgments}
We thank Bai-Cian Ke and Hai-Bo Li for providing us the experimental data.
This work is partly  supported by the National Natural Science Foundation of China
under Grants Nos. 12175066, 11975009.
This work is also supported by the Spanish Ministerio de
Economia y Competitividad (MINECO) and European FEDER funds under Contracts No. FIS2017-84038-C2-1-P
B, PID2020-112777GB-I00, and by Generalitat Valenciana under contract PROMETEO/2020/023. This project has
received funding from the European Union Horizon 2020 research and innovation programme under the program
H2020-INFRAIA-2018-1, grant agreement No. 824093 of the STRONG-2020 project.
This research is also supported by the Munich Institute for Astro-, Particle and BioPhysics (MIAPbP)
which is funded by the Deutsche Forschungsgemeinschaft (DFG, German Research Foundation)
under Germany's Excellence Strategy-EXC-2094 -390783311.


\begin{thebibliography}{}

\bibitem{petrov}  Anders Ryd, Alexey A. Petrov,
Rev. Mod. Phys. \textbf{84} (2012) 65

\bibitem{myreview} E.~Oset, W.~H.~Liang, M.~Bayar, J.~J.~Xie, L.~R.~Dai, M.~Albaladejo, et al.
Int. J. Mod. Phys. E \textbf{25} (2016) 1630001

\bibitem{kaminsky}
J. P. Dedonder, R. Kaminski, L. Lesniak and B. Loiseau, Phys. Rev. D \textbf{89} (2014) 094018

\bibitem{xiedai} J.~J.~Xie, L.~R.~Dai and E.~Oset,
Phys. Lett. B \textbf{742}  (2015) 363

\bibitem{kubis}
F.~Niecknig and B.~Kubis, JHEP \textbf{10}  (2015) 142

\bibitem{toledo}
G.Toledo, N. Ikeno, E. Oset, Eur. Phys. J. C \textbf{81} (2021) 268

\bibitem{mousssallam} E.~Kou, T.~Moskalets and B.~Moussallam,
arXiv:2303.12015 [hep-ph]

\bibitem{newkubis} Franz Niecknig, Bastian Kubis,  Phys. Lett. B \textbf{780} (2018) 471 

\bibitem{enwang} J.~Y.~Wang, M.~Y.~Duan, G.~Y.~Wang, D.~M.~Li, L.~J.~Liu and E.~Wang,
Phys. Lett. B \textbf{821} (2021) 136617

\bibitem{gengxie} X.~Zhu, D.~M.~Li, E.~Wang, L.~S.~Geng and J.~J.~Xie,
Phys. Rev. D \textbf{105} (2022)116010

\bibitem{sunxiao}
Z.~Y.~Wang, J.~Y.~Yi, Z.~F.~Sun and C.~W.~Xiao, Phys. Rev. D \textbf{105} (2022) 016025

\bibitem{wanggeng} X.~Zhu, H.~N.~Wang, D.~M.~Li, E.~Wang, L.~S.~Geng and J.~J.~Xie,
Phys. Rev. D \textbf{107} (2023) 034001

\bibitem{dai}
L.~R.~Dai, E.~Oset and L.~S.~Geng, Eur. Phys. J. C \textbf{82} (2022) 225

\bibitem{patricia} P. C. Magalhaes, M. R. Robilotta, et al.,
Phys. Rev. D \textbf{84} (2011) 094001

\bibitem{luisroca}
L.~Roca and E.~Oset,  Phys. Rev. D \textbf{103} (2021) 034020

\bibitem{focus} J. M. Link \textit{et al.} [FOCUS Collaboration],
Phys. Lett. B \textbf{601} (2004) 10

\bibitem{besexpe}
M.~Ablikim \textit{et al.} [BESIII Collaboration], JHEP \textbf{08} (2022) 196

\bibitem{ollerramos} J.~A.~Oller, E.~Oset and A.~Ramos,
Prog. Part. Nucl. Phys. \textbf{45} (2000) 157

\bibitem{bramon} A. Bramon, A. Grau and G. Pancheri, Phys. Lett. B \textbf{345}  (1995) 263

\bibitem{jiangliang}
J. X. Lin, J. T. Li, S. J. Jiang, W. H. Liang,  E. Oset,
Eur. Phys. J. C \textbf{81} (2021) 1017

\bibitem{pdg}  R. L. Workman et al. (Particle Data Group),
Prog. Theor. Exp. Phys. 2022, 083C01 (2022) and 2023 update

\bibitem{raquel}
R.~Molina, D.~Nicmorus and E.~Oset, Phys. Rev. D \textbf{78} (2008) 114018

\bibitem{geng}
L.~S.~Geng and E.~Oset, Phys. Rev. D \textbf{79}  (2009) 074009

\end{thebibliography}
\end{document}